\hfuzz 2pt
\font\titlefont=cmbx10 scaled\magstep1
\magnification=\magstep1

\def\asymptotic#1\over#2{\mathrel{\mathop{\kern0pt #1}\limits_{#2}}}

\null
\vskip 1.5cm
\centerline{\titlefont MASSLESS NEUTRINO OSCILLATIONS}
\vskip 2.5cm
\centerline{\bf F. Benatti}
\smallskip
\centerline{Dipartimento di Fisica Teorica, Universit\`a di Trieste}
\centerline{Strada Costiera 11, 34014 Trieste, Italy}
\centerline{and}
\centerline{Istituto Nazionale di Fisica Nucleare, Sezione di 
Trieste}
\vskip 1cm
\centerline{\bf R. Floreanini}
\smallskip
\centerline{Istituto Nazionale di Fisica Nucleare, Sezione di 
Trieste}
\centerline{Dipartimento di Fisica Teorica, Universit\`a di Trieste}
\centerline{Strada Costiera 11, 34014 Trieste, Italy}
\vskip 2cm
\centerline{\bf Abstract}
\smallskip
\midinsert
\narrower\narrower\noindent
Quantum dynamical semigroups provide a general framework for studying the
evolution of open systems. Neutrino propagation both in vacuum and 
in matter can be analyzed using these techniques: they allow a consistent
treatment of non-standard, dissipative effects that can alter the
pattern of neutrino oscillations. In particular, initially massless
neutrinos can give rise to a nonvanishing flavour transition probability,
involving in addition the Majorana $CP$-violating mixing phase.
\endinsert
\bigskip
\vfil\eject

{\bf 1. INTRODUCTION}
\medskip

Elementary particle systems are usually treated as isolated quantum systems:
their dynamics can be modeled by means of effective field theories,
allowing a coherent interpretation of the experimental results.
Although very general, this framework can not accommodate all phenomena
involving elementary particles; in particular, those leading to 
irreversibility and dissipation are clearly excluded.
Indeed, a more general treatment is needed to properly describe these effects: 
it can be physically motivated in the framework of open quantum systems.[1-3]

These systems can be thought of as being subsystems in interaction with large
environments. The time evolution of the total system is unitary and
follows the rules of ordinary quantum mechanics; nevertheless,
the dynamics of the subsystem alone, obtained by eliminating 
the environment degrees of freedom, 
shows in general irreversibility and decoherence.

When there are no initial correlations between subsystem and environment
and their mutual interaction is weak, the subdynamics can be described
in a mathematically precise way in terms of quantum dynamical semigroups.
These are linear evolution maps satisfying general properties that
assure the consistent physical interpretation of the dynamics:
they include the condition of entropy increase (irreversibility), 
forward in time composition law (semigroup property), complete positivity.
This framework is very general and can be applied to model irreversibility
and dissipation in very different physical situations;[1-11]
in particular, it can be used to study the evolution 
of elementary particle systems, treated now as open systems.[12-14, 15-19]

The possibility that decoherence phenomena might affect
the physics of elementary particles is supported by recent studies
on the fundamental dynamics of extended objects (strings and
branes);[20] indeed, time evolutions described by quantum dynamical semigroups
can be the result of the interaction with a gas of quanta
obeying infinite statistics ({\it e.g.} a gas of D0-branes).[21]
In other terms, the dynamics of fundamental extended objects 
could effectively generate at low energies a weakly coupled environment.

Similar phenomena have also been described in the framework of quantum
gravity: due to the quantum fluctuation of the gravitational field
and the appearance of virtual black holes, space-time loses its
continuum aspect at distances of the order of Planck's scale and assumes
a foam like behaviour.[22] 
As a consequence, new, non-standard phenomena can arise,
leading to loss of quantum coherence.[23-28]

Unfortunately, our present knowledge of string theory does not allow to 
estimate precisely the magnitude of the non-standard, dissipative 
effects induced on elementary particle systems; they are nevertheless expected
to be very small, being suppressed by at least one inverse power of the
Planck mass, as rough dimensional analysis suggests.
In spite of this, the new effects can affect 
interference phenomena and turn out to be in the reach of future, planned
experiments. Indeed, detailed investigations of neutral meson systems,
neutron interferometry and photon propagation using quantum dynamical
semigroups have already allowed deriving order of magnitude limits on some
of the phenomenological constants parametrizing the new effects,
using available experimental data.[29-31, 17, 19]

In the present work, 
we shall discuss in detail how non-standard, dissipative 
phenomena can affect neutrino propagation, and in  particular neutrino
oscillations. We shall limit our considerations to the
oscillations of two species of neutrinos; in this case, the possible
dissipative effects can be described in terms of six phenomenological
parameters.
A preliminary investigation, limited to vacuum oscillations, has been 
reported in [18]. There, it has been shown that the dissipative phenomena
modify the transition probability $\,\cal P$ among the two 
neutrino flavours, introducing in particular exponential damping factors.
In a simplified situation, limits on one of the dissipative parameters
have subsequently been obtained using recent SuperKamiokande data.[32]

In the following, a much more complete discussion will be presented, with
detailed analysis of oscillation phenomena in presence of irreversibility,
both in vacuum and in matter. Dissipation affects both situations;
in particular, the resonance condition for neutrino propagation in matter
turns out to be modified, leading to distinctive observable effects.
Various approximate expressions for the transition probability $\cal P$
will be given: they can be useful in fitting experimental data.
A discussion on a possible physical mechanism that could give origin to the
non-standard effects will also be presented, although much of the technical
analysis will be relegated to the Appendix.

As a final remark, let us stress that the presence of non-standard, 
dissipative phenomena modify neutrino physics in two important
aspects. First of all, they give the neutrinos an effective mass, 
so that oscillations are possible even for massless neutrinos.
Further, contrary to the standard case, 
the expression of the transition probability $\cal P$ depends in general
on the $CP$-violating phase that is present in the mixing matrix for
Majorana neutrinos.
This allows, at least in principle, 
to distinguish between Dirac {\it vs} Majorana
neutrinos in oscillation experiments. 
We find this possibility as one of the most
intriguing outcome of our investigation.

\vskip 2cm

\line{{\indent\bf 2. NEUTRINOS AS OPEN QUANTUM SYSTEMS}\hfill}
\medskip

The familiar description of neutrino oscillations involves the study
of the evolution of neutrinos created in a given flavour by the weak
interactions and subsequently detected at a later time. The travelling
neutrinos are usually assumed to be ultrarelativistic, so that the
analysis of the transition probability for the original tagged neutrinos
to be found in a different flavour can be performed using an effective 
description.[33-37]

For sake of simplicity, in the following we shall limit our considerations
to the mixing of two neutrino species.%
\footnote{$^\dagger$}{The discussion can be generalized to the case of
three or more neutrinos; however, the explicit formulas for the transition
probabilities would become much more involved and the discussion less
transparent.}
In this case, the neutrino system can be effectively modeled by means of a
two-dimensional Hilbert space; the two neutrino mass eigenstates will
be henceforth fixed as basis in this space. In presence of dissipation,
the physical neutrino states can not be described in terms of elements
of the Hilbert space: a more general formalism is needed
that makes use of density matrices.
These are hermitian, positive operators ({\it i.e.} with non-negative
eigenvalues), normalized to have unit trace.

With respect to the fixed basis, the two flavour states, that we shall
conventionally call $\nu_e$ and $\nu_\mu$, are represented by the
following $2\times 2$ matrices:
$$
\eqalignno{
&\rho_{\nu_e}=\left(\matrix{\cos^2\theta & 
e^{-i\varphi}\cos\theta\,\sin\theta\cr
                   e^{i\varphi}\cos\theta\,\sin\theta & \sin^2\theta}\right)\ ,
&(2.1a)\cr
&\null\cr
&\rho_{\nu_\mu}=\left(\matrix{\sin^2\theta & 
-e^{-i\varphi}\cos\theta\,\sin\theta\cr
              -e^{i\varphi}\cos\theta\,\sin\theta & \cos^2\theta}\right)
\equiv 1-\rho_{\nu_e}\ , &(2.1b)}
$$
where $\theta$ is the ``vacuum'' mixing angle, while the additional phase
$\varphi$ can be nonvanishing for neutrinos of Majorana type.
That this extra phase can not be eliminated by a simple basis redefinition
is a well-known consequence of the reality condition for the Majorana
neutrinos, and, at least in principle, its presence can be experimentally
probed.[38, 39] Nevertheless, in the usual approach, this can not happen via 
the analysis of oscillation phenomena alone.[40] As we shall see, the situation
is different in presence of dissipative effects, so that in the following we
shall keep $\varphi$ nonvanishing unless explicitly stated.

As explained in the introductory remarks, the evolution in time of any neutrino 
state $\rho$ will be described by means of linear maps,
${\mit\Gamma}_t:\rho(0)\mapsto\rho(t)$, that generalize the standard quantum
mechanics unitary evolution. Not all generalized maps $\mit\Gamma_t$
turn out to be physically acceptable: they need to satisfy very general
physical requirements. First of all the maps $\mit\Gamma_t$ should
transform neutrino states into neutrino states, and therefore should map
any initial density matrix $\rho(0)$ into a density matrix
$\rho(t)\equiv{\mit\Gamma}_t[\rho(0)]$, for any $t$. 
Furthermore, they should have
the property of obeying the semigroup composition law,
${\mit\Gamma}_t[\rho(t')]=\rho(t+t')$, for $t,\ t'\geq0$, of increasing
the (von Neumann) entropy, $S=-{\rm Tr}[\rho(t)\ln\rho(t)]$,
of being completely positive.

It has been proved long ago that evolution maps $\mit\Gamma_t$ satisfying
these properties are generated by equations of the following form:[1-3]
$$
{\partial\rho(t)\over \partial t}= -i H_{\rm eff}\ \rho(t)+i\rho(t)\,
H_{\rm eff} + L[\rho(t)]\ .\eqno(2.2a)
$$
The first two pieces in the r.h.s. represent the standard 
quantum mechanical contributions: they give rise to the traditional 
description of neutrino oscillations in terms of the effective (time-independent) 
hamiltonian $H_{\rm eff}$. We shall neglect effects due to possible neutrino 
instability: $H_{\rm eff}$ can then be taken to be hermitian. 
The additional piece $L[\rho]$ is a linear
map that encodes possible dissipative, non-standard effects;
it can be written as:
$$
L[\rho]=-{1\over2}\sum_j\Big(A^\dagger_jA_j\,\rho +
\rho\, A^\dagger_jA_j\Big)\ 
+\sum_j A_j\,\rho\, A^\dagger_j\ ,
\eqno(2.2b)
$$
where the operators $A_j$ must be such that $\sum_j A^\dagger_j A_j$ is a
well-defined $2\times 2$ matrix (entropy increase
can be easily implemented by taking the $A_j$ to be hermitian). 
In absence of it, pure states ({\it i.e.} states of the form 
$|\psi\rangle\langle\psi|$) would be transformed by $\mit\Gamma_t$
into pure states; only when the extra piece $L[\rho]$ is also present,
$\rho(t)$ becomes less ordered in time due to a mixing-enhancing
mechanism: it produces irreversibility and possible loss of quantum
coherence.

In the case of two neutrino flavours, $L[\rho]$ can be fully parametrized
in terms of six, real phenomenological constants,
$a$, $b$, $c$, $\alpha$, $\beta$, and $\gamma$, 
with $a$, $\alpha$ and $\gamma$ non-negative, satisfying the following
inequalities:[1, 15, 16]
$$
\eqalign{
&2\,R\equiv\alpha+\gamma-a\geq0\ ,\cr
&2\,S\equiv a+\gamma-\alpha\geq0\ ,\cr
&2\,T\equiv a+\alpha-\gamma\geq0\ ,\cr
&X\equiv RST-2\, bc\beta-R\beta^2-S c^2-T b^2\geq 0\ .
}\hskip -1cm
\eqalign{
&U\equiv RS-b^2\geq 0\ ,\cr
&V\equiv RT-c^2\geq 0\ ,\cr
&Z\equiv ST-\beta^2\geq 0\ ,\cr
&\phantom{\beta^2}\cr
}\eqno(2.3)
$$
They are direct consequence of the property of complete positivity.
In order for the $2\times 2$ matrix $\rho(t)$ to represent a neutrino
state, its eigenvalues should be positive for any time $t$; this is
crucial for the physical consistency of the whole formalism: the eigenvalues
of $\rho(t)$ are in fact interpreted as probabilities. The property of
complete positivity precisely assures that this holds true in any possible
condition. (For a complete discussion, see [41].)

The one-parameter family of finite evolution $\mit\Gamma_t$ generated by (2.2) 
are called quantum dynamical semigroups; they will be the basis of the
phenomenological treatment of the dissipative effects in the neutrino
system.

The description of irreversible, non-standard phenomena by means of equations
of the form (2.2) is actually very general and can be applied to the
study of very different physical systems. Originally developed in the
framework of quantum optics,[5-7] 
it has also been successfully used in the analysis
of statistical models,[1-3] the interaction of a microsystem with a 
measuring apparatus,[8-11] the study of dissipative effects in systems involving
elementary particles, in particular neutral mesons.[15, 16, 29-31] 
Although essentially phenomenological in nature, 
all these analysis can be supported by physical considerations.

A general picture in which the quantum dynamical semigroup description of
dissipative effects naturally emerges is provided by open systems,
{\it i.e.} by systems in weak interactions with a large environment.
In the case of elementary particles, 
these effects are likely to originate from the fundamental
dynamics of strings or branes, which is in general rather complex. 
Nevertheless, an effective description of the
environment that encodes some of the properties of the underlying fundamental
dynamics turns out to be adequate for a more physical discussion of evolution
equations of type (2.2).

Quite in general, the total hamiltonian of a system $\cal S$ in interaction
with an environment $\cal E$ can be decomposed as
$$
H_{\rm tot}=H\otimes {\bf 1} + {\bf 1}\otimes H_{\cal E} + 
g\,H'\ ,
\eqno(2.4)
$$
where $H$ is the system hamiltonian in absence of $\cal E$,
while $H_{\cal E}$ drives the internal dynamics of the environment. 
The interaction between $\cal S$ and $\cal E$ is described by $H'$,
with $g$ a small, dimensionless coupling constant.

In many instances, the initial state of the total system
${\cal S} + {\cal E}$ can be taken to be in factorized form:
$\rho_{\rm tot}=\rho\otimes\rho_{\cal E}$. This is surely
justified in the case of the neutrino system: since the mechanism of
neutrino production is different from the one responsible for the
dissipative effects, system and environment are surely uncorrelated
at the moment of the emission.%
\footnote{$^\dagger$}{Even in presence of an initially correlated total
system ${\cal S} + {\cal E}$, the factorized approximation 
becomes a very good approximation when the short-time correlations 
have died out.[4]}
Then, the time evolution of the state $\rho$ of the system $\cal S$
can be obtained by tracing over the environment degrees of freedom:
\null
$$
\rho\equiv\rho(0)\mapsto\rho(t)=
{\rm Tr}_{\cal E}\Big[ e^{-iH_{\rm tot} t}\,
\big(\rho\otimes\rho_{\cal E}\big)\, e^{iH_{\rm tot} t}\Big]\ ,
\eqno(2.5)
$$
\null

In general, the resulting map $\rho(0)\mapsto\rho(t)$ turns out to be 
rather involved, developing non-linearity and memory effects. Nevertheless,
when the interaction between $\cal S$ and $\cal E$ is weak, an evolution
equation for $\rho(t)$ local in time naturally emerges. 
The technical details are presented in the Appendix. As discussed there,
the environment can be modeled as a gas of quanta, obeying infinite statistics; 
this description is in line with the idea that the dissipative
effects originate from the low energy string dynamics at a fundamental scale
$M_F$ ({\it e.g.} Planck's mass). Then, in the weak coupling limit,
{\it i.e.} when the coupling constant $g$ becomes very small,
the resulting dynamical equation for the subsystem state $\rho(t)$
turns out to be precisely of the form (2.2).[1-3, 21]

This result allows a rough estimate of the magnitude
of the effects produced by the non-standard piece $L[\rho]$: they should be
proportional to powers of the typical energy of the system $\cal S$,
while suppressed by inverse powers of the characteristic energy scale
of $\cal E$. In the case of the neutrino system, these effects should
be very small, since the typical energy scale of the environment
can be assimilated to the fundamental scale $M_F$. For any fixed
neutrino source and observational conditions, an upper bound on the
magnitude of the effects induced by $L[\rho]$ can be evaluated to be of
order $E^2/M_F$, where $E$ is the average neutrino energy.

As a further outcome of the weak coupling limit procedure, the
hamiltonian part of the evolution equation for $\rho(t)$ gets
modified by the presence of the environment. Indeed, the effective
hamiltonian $H_{\rm eff}$ in (2.2) does not coincide in general with the
starting system hamiltonian $H$ in (2.4): suitable dissipative
contributions to $H$, generated by the interaction $H'$, need to be
taken into account.[1-3, 21] 
As we shall see in the following, this fact has interesting consequences 
in neutrino physics: one can have oscillations among different flavours
induced by dissipative effects even for massless (or mass-degenerate)
neutrinos. In other words, originally massless neutrinos can get an 
effective non-zero mass via the interaction with the environment.

\vfill\eject

\line{{\indent\bf 3. QUANTUM DYNAMICAL SEMIGROUPS AND NEUTRINO}\hfill}
\line{\phantom{\bf\indent 3. }{\bf OSCILLATIONS}\hfill}
\medskip

In the case of the neutrino system, much of the considerations
and discussions of the previous section about the evolution equation (2.2)
can be made more transparent and explicit. In particular, both for
the effective hamiltonian $H_{\rm eff}$ and for the extra piece $L[\rho]$
simple expressions can be given. 

We shall be as general as possible and
include in our discussion effects due to the propagation of neutrinos
in a medium made of ordinary matter. Because of the interactions of
the neutrinos with the particles in the medium, an effective potential
can be generated, that has different effects for different flavours.
In the case of ordinary matter, the electron neutrinos interact with
the electrons in the medium, so that their average energy effectively receive
an extra contribution $A=\sqrt{2}\, G_F\, n_e$ with respect to the
energy of the muon neutrinos ($G_F$ is the Fermi constant, while $n_e$
represents the electron number density in the medium).[42, 33-37]
In the ordinary case, this contribution can significally change the
oscillation pattern between $\nu_e$ and $\nu_\mu$ states (the so-called
MSW effect).[43, 44] 
As we shall see, this phenomenon can be substantially modified
by the presence of non-standard, dissipative effects.

In the basis introduced in the previous section, the $2\times 2$ matrix 
representing the effective hamiltonian can be taken to be of the form:
$$
H_{\rm eff}=\left(\matrix{
E-\omega_0-\omega_3&\omega_1-i\omega_2\cr
\omega_1+i\omega_2&E+\omega_0+\omega_3\cr}\right)+
{A\over2}\left(\matrix{1+\cos2\theta & e^{-i\varphi}\sin2\theta\cr
e^{i\varphi}\sin2\theta & 1-\cos2\theta}\right)
\ . \eqno(3.1)
$$
In the first piece, $E$ represent the average neutrino energy, while
$\omega_0=\Delta m^2/4E$ takes into account the square mass difference
$\Delta m^2$ of the two mass eigenstates; these are the usual contributions
that give rise to the standard oscillation pattern in vacuum.
The extra real parameters $\omega_1$, $\omega_2$ and $\omega_3$ are 
the consequence of the interaction with the environment; as explained
in the previous section (and discussed in detail in the Appendix),
they represent the contribution of the dissipative phenomena
to the system hamiltonian.

Both $\omega_0$ and $\omega_1$, $\omega_2$, $\omega_3$ contribute
to the level splitting 
$\omega=[(\omega_0+\omega_3)^2+\omega_1^2+\omega_2^2]^{1/2}$
between the two mass eigenstates, so that they all contribute
to the oscillation phenomena in vacuum. Therefore, even for initially
degenerate mass eigenstates, $\Delta m^2=\,0$, vacuum oscillations
can occur between the two flavours due to the dissipative effects
induced by the fundamental dynamics at the large scale $M_F$.
Although in general all three parameters $\omega_1$, $\omega_2$ 
and $\omega_3$ are non-vanishing, in the following, in order to
simplify the treatment, we shall assume
$\omega_1=\omega_2=\,0$; this working assumption allows for more
manageable formulas, while keeping unaffected their physical meaning
and implications.%
\footnote{$^\dagger$}{When $\Delta m^2=\,0$, this is no longer an
assumption: in this case, one can always choose to work in a basis
for which $\omega_1$ and $\omega_2$ vanish.}

The final contribution to $H_{\rm eff}$ in (3.1) 
takes into account the interaction of the
propagating neutrinos with ordinary matter; it would be diagonal in the
flavour basis (only electron neutrinos are affected), but assumes
a more complicated expression involving the mixing angle $\theta$ and the phase
$\varphi$ in the chosen basis. Since the coefficient $A$ is proportional
to the density of electrons in the mean, for propagation in 
non-homogeneous matter $H_{\rm eff}$ will in general be a function of
the position of the neutrinos. Nevertheless, one can always approximate
a non-homogeneous medium by a collection of media, 
each with a constant density, while having different thickness; 
in view of this, in the following we shall assume
the parameter $A$ to be a constant (see also the discussion in Sect.4).

As mentioned before, although the effective hamiltonian
$H_{\rm eff}$ gets also dissipative contributions, only when the
additional piece $L[\rho]$ in the evolution equation (2.2) is
non-vanishing, irreversibility and mixing enhancing effects are possible.
In the present case, its explicit expression in terms of the six
phenomenological constants $a$, $b$, $c$, $\alpha$, $\beta$, and $\gamma$
in (2.3) can be most simply given by expanding the $2\times 2$
matrix $\rho$ in terms of the Pauli matrices $\sigma_i$, $i=1,2,3$,
and the identity $\sigma_0$:
$$
\rho={1\over 2}\sum_{\mu=0}^3\rho_\mu\, \sigma_\mu\ .
\eqno(3.2)
$$
In this way, the linear map $L$ acting on $\rho$ can be represented by
the following, symmetric $4\times 4$ matrix $\big[L_{\mu\nu}\big]$, 
acting on the 4-vector of components $(\rho_0,\rho_1,\rho_2,\rho_3)$:
$$
\big[L_{\mu\nu}\big]=-2\left(\matrix{0&0&0&0\cr
                                     0&a&b&c\cr
                                     0&b&\alpha&\beta\cr
                                     0&c&\beta&\gamma\cr}\right)
\ .\eqno(3.3)
$$

The form of the evolution equation (2.2) can be further simplified by
recalling that it is trace preserving. From the initial normalization
condition ${\rm Tr}[\rho(0)]=1$, one immediately obtains that the component
of $\rho(t)$ along the identity is equal to one for all times.
Then, the evolution equation for the remaining three components of $\rho(t)$
can be rewritten in a Scr\"odinger-like form:
$$
{\partial\over\partial t}|\rho(t)\rangle =-2\, {\cal H}\ |\rho(t)\rangle\ ,
\eqno(3.4)
$$
where the 3-vector $|\rho\rangle$ has components
$(\rho_1,\rho_2,\rho_3)$, while
$$
{\cal H}=\left(\matrix{a & b+\mu & c-\nu\sin\varphi\cr
                       b-\mu & \alpha & \beta+\nu\cos\varphi\cr
                       c+\nu\sin\varphi & \beta-\nu\cos\varphi & \gamma}\right)
\ ,\eqno(3.5)
$$
with
$$
\mu={A\over 2}\cos2\theta-\omega\ ,\qquad\qquad \nu={A\over2}\sin2\theta\ .
\eqno(3.6)
$$
The solution of (3.4) involves the formal exponentiation of the
matrix $\cal H$:
$$
|\rho(t)\rangle= {\cal M}(t)\ |\rho(0)\rangle\ ,\qquad
{\cal M}(t)=e^{-2\,{\cal H}\,t}\ .\eqno(3.7)
$$

As discussed in [18], expressions for the entries of ${\cal M}(t)$ can
always be obtained by solving the eigenvalue problem for
the $3\times 3$ matrix in (3.5):
$$
{\cal H}\, |v^{(k)}\rangle =\lambda^{(k)}\, |v^{(k)}\rangle\ ,
\qquad k=1,2,3\ .\eqno(3.8)
$$
The three eigenvalues $\lambda^{(1)}$, $\lambda^{(2)}$, $\lambda^{(3)}$
satisfy the cubic equation:
$$
\lambda^3+r\, \lambda^2+ s\, \lambda +w=\,0\ ,\eqno(3.9)
$$
with real coefficients:

$$
\eqalignno{
&r\equiv -(\lambda^{(1)}+\lambda^{(2)}+\lambda^{(3)})=
-(a+\alpha+\gamma)\ , &(3.10a)\cr
&\null\cr
&s\equiv \lambda^{(1)}\lambda^{(2)}+\lambda^{(1)}\lambda^{(3)}+
\lambda^{(2)}\lambda^{(3)}=a\alpha + a\gamma + \alpha\gamma
-b^2-c^2-\beta^2 +\mu^2+\nu^2\ ,\qquad &(3.10b)\cr
&\null\cr
&w\equiv -\lambda^{(1)}\lambda^{(2)}\lambda^{(3)}=
a(\beta^2-\nu^2\cos^2\varphi)+\alpha(c^2-\nu^2\sin^2\varphi)
+\gamma(b^2-\mu^2)\cr
&\hskip 3cm -a\alpha\gamma-2\,bc\beta-b\nu^2\sin2\varphi
-2\mu\nu(\beta\sin\varphi+c\cos\varphi)\ .
&(3.10c)\cr}
$$

\noindent
The solutions are either real, or one is real and the remaining two are
complex conjugate, according to the sign of the associated discriminant:
$D=p^3+q^2$, $p=s/3-(r/3)^2$, $q=(r/3)^3-rs/6+w/2$ (degenerate,
real solutions occur when $D=\,0$).[45]
Then, recalling that the matrix $\cal H$ itself satisfy the
equation (3.9), one can derive the following expression for the entries 
of ${\cal M}(t)$:
$$
{\cal M}_{ij}(t)=\sum_{k=1}^3 e^{-2\lambda^{(k)} t}
\left[
{\big([\lambda^{(k)}]^2+r\lambda^{(k)}+s\big)\delta_{ij}
+\big(\lambda^{(k)}+r\big){\cal H}_{ij}+{\cal H}^2{}_{ij}\over
3[\lambda^{(k)}]^2+2r\lambda^{(k)}+s}\right]\ ,\quad i,j=1,2,3\ .
\eqno(3.11)
$$

Although rather formal, this formula allows a general discussion on the
behaviour of ${\cal M}(t)$. For $\mu=\nu=\,0$, thanks to the inequalities
in (2.3), the matrix $\cal H$ results real, symmetric and non-negative:
its eigenvalues are all real and non-negative. Only when $|\mu|$ and $|\nu|$
are sufficiently large, complex eigenvalues may appear, although with
a non-negative real part, since in general the evolution generated by (2.2)
is bounded for any $t$.[46] In this case an oscillatory behaviour is
possible, while for small $\mu$, $\nu$, the damping terms prevail and 
dissipation is the dominant phenomena.

In particular, since generically ${\rm det}{\cal H}\equiv-w\neq 0$,
in presence of dissipation the real part of
$\lambda^{(1)}$, $\lambda^{(2)}$, $\lambda^{(3)}$ are all strictly positive;
therefore ${\cal M}(t)$ asymptotically vanish for large enough times.%
\footnote{$^\dagger$}{In presence of vanishing eigenvalues, this decoherence
effect is only partial;[18] however, note that having
${\rm det}{\cal H}=\,0$ requires a unnatural fine-tuning among the parameters
in $(3.10c)$.}
This has clearly dramatic consequences in the study of neutrino flavour
transitions.

Let us assume that at $t=\,0$ the neutrinos are generated to be of type
$\nu_e$. In the formalism of density matrices, the probability of
having a transition into neutrinos of type $\nu_\mu$ at time $t$
is given by:
$$
{\cal P}_{\nu_e\to\nu_\mu}(t)\equiv{\rm Tr}
\Big[\rho_{\nu_e}(t)\ \rho_{\nu_\mu}\Big]
={1\over 2}\bigg[1+\sum_{i,j=1}^3
\rho_{\nu_\mu}^i\, \rho_{\nu_e}^j\ {\cal M}_{ij}(t)\bigg]\ ,
\eqno(3.12)
$$
where $\rho_{\nu_e}(t)$ is the solution of (2.2) with the initial condition
given by the matrix $\rho_{\nu_e}(0)\equiv\rho_{\nu_e}$, while
$\rho_{\nu_e}^i$, $\rho_{\nu_\mu}^j$, $i,j=1,2,3$ are the components
of the 3-vectors $|\rho_{\nu_e}\rangle$, $|\rho_{\nu_\mu}\rangle$
corresponding to the density matrices in (2.1).
Using the explicit expressions for these components, one finds:
$$
\eqalign{
{\cal P}_{\nu_e\to\nu_\mu}(t)=&{1\over2}\bigg\{1-
\cos^22\theta\,{\cal M}_{33}(t)\cr
&-\sin^2 2\theta\, \Big[{\cal M}_{11}(t)\, \cos^2\varphi +
{\cal M}_{22}(t)\, \sin^2\varphi +
\Big({\cal M}_{12}(t)+{\cal M}_{21}(t)\Big)\, \sin\varphi\cos\varphi\Big]\cr
&-\cos2\theta\, \sin2\theta\, \Big[
\Big({\cal M}_{13}(t)+{\cal M}_{31}(t)\Big)\ \cos\varphi
+\Big({\cal M}_{23}(t)+{\cal M}_{32}(t)\Big)\ \sin\varphi\Big]\bigg\}\ .
}
\eqno(3.13)
$$
\null

One of the interesting features of this formula 
is its explicit dependence on the
phase $\varphi$; in presence of dissipative effects, it is therefore
possible, at least in principle, to distinguish between Dirac and Majorana
neutrinos by studying the oscillation pattern in (3.13). This peculiarity
disappears when the non-standard, dissipative pieces in (2.2) are absent;
indeed, in that case, one has:
$$
{\cal M}_{ij}(t)=\delta_{ij}-{\sin2\omega_M t\over\omega_M}\, {\cal H}_{ij}
+{2\sin^2\omega_M t\over\omega_M^2}\, {\cal H}^2{}_{ij}
\qquad i,j=1,2,3\ ,
\eqno(3.14)
$$
where $\cal H$ is now as in (3.5) with $a$, $b$, $c$, $\alpha$, 
$\beta$, and $\gamma$ all equal to zero, while
$\omega_M=\sqrt{\mu^2+\nu^2}$,
and (3.13) reduces to the well-known standard expression
for the oscillation probability in an homogeneous medium:[42, 33-37]
$$
{\cal P}^{(0)}_{\nu_e\to\nu_\mu}(t)=\sin^2 2\theta_M\ \sin^2\omega_M t\ ,
\qquad\quad 
\sin^2 2\theta_M={\sin^2 2\theta\over \Big({A\over2\omega} -\cos2\theta\Big)^2+
\sin^2 2\theta}\ .
\eqno(3.15)
$$

Another distinctive characteristic of the transition probability
in presence of dissipation given in (3.13) is its 
asymptotic behaviour for large times, 
which turns out to be independent from the
mixing angle $\theta$, the phase $\varphi$ and the matter coefficient $A$:
$$
{\cal P}_{\nu_e\to\nu_\mu}(t)\ \asymptotic\sim\over{t\to\infty}\
{1\over2}\ .\eqno(3.16)
$$
This result is a direct consequence of the vanishing of the 
matrix ${\cal M}(t)$ in (3.7).
Nevertheless, as discussed below, the regime of validity of this asymptotic
limit can seldom be reached in practical experimental conditions.

\vfill\eject

{\bf 4. TRANSITION PROBABILITY IN MATTER}
\medskip

The general expression (3.13) for the transition probability
is rather involved and it is not
particularly useful for studying in more detail its physical
properties. Therefore, in the present and following sections we shall
discuss various approximations in which 
${\cal P}_{\nu_e\to\nu_\mu}(t)$ assumes a more manageable form.
These simplified expressions, besides being appropriate
for theoretical analysis, could also be used to fit
actual experimental data.[32, 47]

As already mentioned in Sect.2, the values of the constants
$a$, $b$, $c$, $\alpha$, $\beta$, and $\gamma$ parametrizing the
non-standard effects, are expected 
to be very small, with an upper bound of order
$E^2/M_F\simeq 10^{-19}\ {\rm GeV}$, for $E\simeq 1\ {\rm GeV}$
and for $M_F$ the Planck mass. Nevertheless, this estimate is not far from
the values that the standard oscillation parameter
$\omega_0=\Delta m^2/4E$ assumes for typical neutrino sources. Indeed,
the ratio of $a$, $b$, $c$, $\alpha$, $\beta$, and $\gamma$ with
$\omega_0$ can be evaluated to be at most of order
$10^{-10} E^3/\Delta m^2$, with $E$ expressed in MeV and the neutrino
mass difference $\Delta m^2$ in ${\rm eV}^2$; this ratio turns out to be
about $10^2$ for atmospheric neutrinos, 
of order one for solar neutrinos,
while for accelerator neutrinos it can be as small as $10^{-2}$.
Therefore, the effects induced by dissipation can interfere with those
producing oscillations via a non-vanishing $\omega_0$, resulting in
observable modifications of the oscillation pattern.
Present and, most likely, future dedicated neutrino experiments
should be able to detect these modifications, or at least put 
stringent limits on the magnitude of the non-standard phenomena.

Let us first consider the case in which the dissipative
parameters $a$, $b$, $c$, $\alpha$, $\beta$, and $\gamma$
are of the same order or larger than the remaining constants in (3.1).
In this case a very useful approximation is to assume $a=\alpha=\gamma$
and $c=\,0$, conditions perfectly compatible with the inequalities
(2.3), provided $\alpha^2\geq b^2+\beta^2$. For simplicity,
we further assume the extra phase $\varphi$ to be vanishingly small.
A manageable expression for the transition probability can then be derived:
$$
{\cal P}_{\nu_e\to\nu_\mu}(t)={1\over2}\Big(1-e^{-2\alpha t}\Big)
+\bigg[{ {\tilde\nu}^2-{\tilde\beta}^2\over\Omega_M^2}\bigg]\,
e^{-2\alpha t}\, \sin^2(\Omega_M t)\ ,
\eqno(4.1)
$$
where
$$
\Omega_M=\big[\mu^2+\nu^2-b^2-\beta^2\big]^{1/2}\ ,
\eqno(4.2)
$$
$$
\tilde\nu=\omega\,\sin 2\theta\ ,\qquad\quad
\tilde\beta=\beta\, \cos 2\theta+b\, \sin 2\theta\ .
\eqno(4.3)
$$
The oscillating behaviour in (4.1) depends on the magnitude of the combination
$\mu^2+\nu^2=(A/2-\omega\,\cos 2\theta)^2+\omega^2\, \sin^2 2\theta$
with respect to $b^2+\beta^2$; in regions for which
$b^2+\beta^2\geq \mu^2+\nu^2$, 
the frequency $\Omega_M$ becomes purely imaginary
and ${\cal P}_{\nu_e\to\nu_\mu}(t)$ contains only exponential terms.
Anyway, the $\alpha$-depending damping terms in (4.1)  
dominate for large times, and the asymptotic limit (3.16) is thus recovered.

In absence of dissipation, $\alpha=b=\beta=\,0$, the factor
in front of the sine term in (4.1) becomes parametrized as
in (3.15), with a modified mixing angle $\theta_M$. 
When the matter parameter $A$
is close to $A_R\equiv 2\omega\cos 2\theta$, the transition probability
gets enhanced, and oscillations between the two neutrino species is possible
even when the original mixing angle $\theta$ is small. This phenomenon
is at the root of the so-called MSW effect.[43, 44, 42]

In presence of dissipation however, the physical consequences of this effect
are in general much more modest. In this case, one can not parametrize
${\cal P}_{\nu_e\to\nu_\mu}(t)$ in terms of a modified mixing angle;
in spite of this, the expression in (4.1) as a function of $A$, at fixed time,
has a critical point for $A=A_R$. This point is a maximum for
${\tilde\nu}^2\geq {\tilde\beta}^2$, and indeed as $A$ approaches $A_R$
an enhancement in ${\cal P}_{\nu_e\to\nu_\mu}(t)$ occurs:
${(\tilde\nu}^2-{\tilde\beta}^2)/\Omega_M^2>1$;
however, the exponentially damping factors in (4.1) greatly reduce
in practice its effectiveness.
Further, when ${\tilde\nu}^2< {\tilde\beta}^2$, the probability
${\cal P}_{\nu_e\to\nu_\mu}(t)$ in (4.1) is maximally suppressed at
the critical point: it is dominated by the damping factors.

This discussion might appear spoiled by the initial assumption of a constant
matter parameter $A$: the occurrence of the MSW effect requires
a medium with a (slowly) varying density. As already pointed out,
the assumption of a constant $A$ is not really a limitation: one can
always approximate, with arbitrary accuracy, the travelling of neutrinos
through varying density matter as the propagation in a series of media
with different constant densities and different thickness.
The total time-evolution will be given by the composition of the
evolutions in the various matter slices, so that the matrix
${\cal M}$ in (3.7) becomes:
$$
{\cal M}(t)={\cal M}_n(t_n)\cdots{\cal M}_2(t_2)\,{\cal M}_1(t_1)\ ,
\qquad t=t_1+t_2+\ldots+t_n\ ,
\eqno(4.4)
$$
where, $t_1,\ t_2,\ldots,t_n$ are the total times spent by the neutrinos
in the various media, while ${\cal M}_i$, $i=1,\ldots,n$ are the corresponding
propagation matrices.

As an example, let us consider the case of an initial electron neutrino
travelling for a time $t_1$ into a medium with matter parameter $A$, that is 
then detected in vacuum at a later time $t=t_1+t_2$; this situation can
roughly represent a solar neutrino model.
Using (4.4), the probability of detecting
the original $\nu_e$ as a muon neutrino is given by:
$$
\eqalign{
{\cal P}_{\nu_e\to\nu_\mu}(t)=&{1\over2}\bigg\{1-
e^{-\alpha(t_1+t_2)}\cr
-&e^{-\alpha(t_1+t_2)}
\bigg({ {\tilde\nu}^2-{\tilde\beta}^2\over\Omega_0^2}\bigg)
\bigg[{4\Omega_M^2-A(A+2b+2\omega\cos 2\theta)\over\Omega_M^2}
\sin^2(\Omega_M t_1)\,\sin^2(\Omega_0 t_2)\cr
-&{\Omega_0\over\Omega_M}\, \sin(2\Omega_M t_1)\ \sin(2\Omega_0 t_2)
-2\sin^2(\Omega_0 t_2)
-2{\Omega_0^2\over\Omega_M^2}\ \sin^2(\Omega_M t_1)\bigg]\bigg\}\ ,
}
\eqno(4.5)
$$
where $\Omega_0=\sqrt{\omega^2-b^2-\beta^2}$, while $\Omega_M$ is as
in (4.2). One can check that, with the appropriate choice of parameters
(see the previous discussion), the probability 
${\cal P}_{\nu_e\to\nu_\mu}(t)$ can indeed get an enhancement
for $A$ close to the critical point; the effect is however modest
and further suppressed by the damping factors. Nevertheless,
with the appropriate values of $A$, $t_1$ and $t_2$,
the expression (4.5) can be used to fit solar neutrino data;
the total flight time $t$ is large, so that a good sensitivity
at least on the dissipative parameter $\alpha$ is surely
attainable.

A different approximation of the full expression (3.13) for the
transition probability ${\cal P}_{\nu_e\to\nu_\mu}$ can be
obtained when the dissipative parameters
$a$, $b$, $c$, $\alpha$, $\beta$, and $\gamma$, can be considered
small with respect to the level-splitting term $\omega$; as mentioned
before, this typically occur for neutrino beams generated at accelerators.
In this case the additional term $L[\rho]$ in the evolution equation (2.2)
can be treated as a perturbation. To first order in the small
parameters, explicit expressions for the entries of the evolution matrix
${\cal M}(t)$ in (3.7) can be easily obtained; then, using (3.12),
one finds:
$$
{\cal P}_{\nu_e\to\nu_\mu}(t)=
\bigg({{\tilde\mu}^2\over\omega^2_M}\bigg)\, 
e^{-2\lambda_1 t}+e^{-\lambda_2 t}
\bigg[\bigg({{\tilde\nu}^2\over\omega_M^2}\bigg)\,
\cos(2\,\omega_M t) + \bigg({N\over\omega_M^2}\bigg)\, 
\sin(2\,\omega_M t)\bigg]\ ,
\eqno(4.6)
$$
where $\omega_M=\sqrt{\mu^2+\nu^2}$ as in the previous section, while
$$
\eqalignno{
&\tilde\mu=\mu\cos 2\theta+\nu \sin 2\theta={A\over 2}-\omega\cos 2\theta
\ ,&(4.7a)\cr
&\tilde\nu=\nu\cos 2\theta-\mu \sin 2\theta=\omega \sin 2\theta\ ; &(4.7b)}
$$
the parameters $\lambda_1$, $\lambda_2$ and $N$ contains the dependence
on the dissipative constants:
$$
\eqalignno{
&\lambda_1=\big(a\nu^2+2c\mu\nu+\gamma\mu^2\big)/\omega_M^2\ , &(4.8a)\cr
&\lambda_2=\alpha+\big(a\mu^2-2c\mu\nu+\gamma\nu^2\big)\omega_M^2\ ,&(4.8b)\cr
&N={{\tilde\nu}^2\over 2}\Big[\alpha-a-\nu(3a\nu+2c\mu)/\omega_M^2\Big]
+3\nu{\tilde\nu}(a-\gamma)\cos 2\theta \cr
&\hskip 6cm+c\Big(2\mu{\tilde\mu}^2\nu-\omega_M^4\,
\sin 4\theta\Big)/\omega_M^2\ .\qquad&(4.8c)}
$$
In the expression (4.6), we have reconstructed the exponential factors
by consistently putting together the terms linear in $t$.
Notice that the result in (4.8) is in
agreement with the discussion in Sect.3 concerning the eigenvalues
of the matrix $\cal H$. In this case the algebraic equation (3.9) 
has one real, $\lambda^{(1)}$, and two complex conjugate solutions, 
$\lambda^{(2,3)}=\lambda_R\pm i\lambda_I$; within our approximation,
$\lambda^{(1)}=\lambda_1$, $2\lambda_R=\lambda_2$, so that the first condition
in (3.10) is satisfied, while the remaining two fix the imaginary 
part $\lambda_I$.

The expression (4.6) for the transition probability can be used to
fit experimental data. With respect to the standard case, it contains
three additional parameters, $\lambda_1$, $\lambda_2$ and $N$,
that signal the presence of non-standard phenomena, through
the constants $a$, $b$, $c$, $\alpha$, $\beta$, and $\gamma$.
If at least one of these three parameters is found to be non zero,
it would clearly signal the presence of dissipative effects in neutrino
physics; this is surely the most simple experimental check on the
generalized evolution equation (2.2) that can be performed
with accelerator neutrino beams.

\vskip 2cm

{\bf 5. ADIABATIC APPROXIMATION}
\medskip

For time evolutions with a semigroup composition law, the appropriate
way to follow neutrino propagation in a medium is through the successive
applications of the finite evolution matrices ${\cal M}(t)$ to the
initial state $|\rho(0)\rangle$, as shown in (4.4). In more traditional
approaches, one usually adopts a different approximation, based
on the assumption of (adiabatic) slowly varying matter density.
This approximation can be easily discussed also in the framework
of density matrices and quantum dynamical semigroups.

Let us consider the case of a neutrino, created at $t=\,0$ in matter
of high density ({\it i.e.} in the core of the sun), propagating
towards regions of smaller density. In this case, the effective
hamiltonian (3.1) is no longer constant,
and the propagating matrix ${\cal M}(t)$ involves a time-ordered
exponentiation of $\cal H$. Nevertheless, at any instant of time, the
$3\times 3$ matrix $\cal H$ can be diagonalized by a similarity 
transformation:
$$
{\cal H}=T\, {\cal D} \, T^{-1}\ .
\eqno(5.1)
$$
Using this decomposition in (3.4), one can derive
the evolution equation for the transformed 3-vector, 
$|\tilde\rho\rangle= T\, |\rho\rangle$; explicitly, one finds
$$
{\partial\over\partial t} |\tilde\rho(t)\rangle=
-2\bigg[{\cal D}+{\partial T\over\partial t}\, T^{-1}\bigg]\ 
|\tilde\rho(t)\rangle\ .
\eqno(5.2)
$$
The adiabatic approximation amounts to neglecting the last term
in this equation; this is justified when the matter density parameter $A$
is slowly varying. In this approximation, the neutrino state essentially 
evolves in time as an eigenstate of $\cal H$.

In order to make the discussion more explicit, as in the previous section
we shall take $a=\alpha=\gamma$ and $c=\,0$, while neglecting the
extra phase $\varphi$ and the dissipative contributions $\omega_1$,
$\omega_2$ to $\cal H$. With these choices, one has:
$$
{\cal D}=-2\,\left(\matrix{\alpha &  &  \cr
                        & \alpha +i\Omega_M & \cr
					    &  & \alpha-i\Omega_M}\right)\ ,
\eqno(5.3)
$$
with $\Omega_M$ as in (4.2), while the transformation matrix $T$ 
takes the form:
$$
T={1\over\sqrt{2}\Omega_M}\left(\matrix{\sqrt{2}\, (\beta+\nu) & 
b+\mu & b+\mu \cr
0 & i\Omega_M & -i\Omega_M \cr
-\sqrt{2}\, (b-\mu) & \beta-\nu & \beta-\nu }\right)\ .
\eqno(5.4)
$$
The entries of $T$ can be parametrized in terms of two real variables 
$\xi$ and $\zeta$ and an angle $\phi$, that could be complex:
$$
\eqalignno{
&{\beta\pm\nu\over\Omega_M}=\pm\Big(e^{\pm\zeta}\cos 2\theta\sin 2\phi
+ e^{\pm\xi} \sin2\theta\cos2\phi\Big)\ , &(5.5a)\cr
&{b\pm\mu\over\Omega_M}=\pm\Big(e^{\mp\zeta}\sin2\theta\sin2\phi
-e^{\mp\xi}\cos 2\theta\cos2\phi\Big)\ ; &(5.5b)\cr
}
$$
for later convenience, an explicit dependence on the mixing angle $\theta$
has been extracted from the entries of $T$.

Using a compact vector notation, the transition probability
${\cal P}_{\nu_e\to\nu_\mu}$ in (3.12) can now be written as
$$
{\cal P}_{\nu_e\to\nu_\mu}(t)={1\over 2}\Big\{ 1
+\langle\rho_{\nu_\mu}|\ T_f\cdot {\cal M}_{\cal D}(t_f,t_i)\cdot T_i^{-1}\
|\rho_{\nu_e}\rangle\Big\}\ ,
\eqno(5.6)
$$
where
$$
{\cal M}_{\cal D}(t_f,t_i)=e^{-2\alpha t}\
\left(\matrix{1 &  & \cr
 & e^{-2i\int_0^t d\tau \Omega_M(\tau)} & \cr
 &  & e^{2i\int_0^t d\tau \Omega_M(\tau)}}\right)\ ,
\eqno(5.7)
$$
while $T_i$ and $T_f$ are the matrices that diagonalize $\cal H$ at the
initial time $t_i=\,0$ and final time $t_f=t$; they can be written
as in (5.4), with parameters $\xi_i$, $\zeta_i$, $\phi_i$ 
and $\xi_f$, $\zeta_f$, $\phi_f$, respectively. Explicitly, one finds
$$
\eqalign{
{\cal P}_{\nu_e\to\nu_\mu}(t)={1\over 2}\bigg\{1
&-e^{-2\alpha t}\Big[e^{\xi_f-\xi_i}\, \cos 2\phi_f\, \cos 2\phi_i\cr
&+ e^{-(\zeta_f-\zeta_i)}\, \sin 2\phi_f\, \sin 2\phi_i\
\cos\Big(2\int_0^t d\tau\, \Omega_M(\tau)\Big)\Big]\bigg\}\ .
}
\eqno(5.8)
$$
In absence of dissipation, $\xi_i=\zeta_i=\xi_f=\zeta_f=\, 0$, 
$\alpha=b=\beta=\, 0$,
one recovers the familiar expression for the adiabatic transition
probability.

When the adiabatic approximation ceases 
to be valid, the previous treatment
needs to be generalized. Indeed, in this case, the neutrino state
no longer remains in a specific eigenstate of $\cal H$ for the whole
time evolution; rather, it can mix with the remaining eigenstates.
In order to take into account this possibility, the expression for
the transition probability (5.6) needs to be modified:
$$
{\cal P}_{\nu_e\to\nu_\mu}(t)={1\over 2}\Big\{ 1
+\langle\rho_{\nu_\mu}|\ T_f\cdot {\cal M}_{\cal D}(t_f,t_c)\cdot
{\mit\Delta}\cdot {\cal M}_{\cal D}(t_c,t_i)\cdot T_i^{-1}\
|\rho_{\nu_e}\rangle\Big\}\ ,
\eqno(5.9)
$$
where $t_c$ is the time at which the neutrino crosses the critical region,
while $\mit\Delta$ is the mixing matrix that encodes the possible
hopping between the instantaneous eigenstates of the effective hamiltonian.
For simplicity, we are not taking into account the possibility of
hoppings induced by dissipative effects: they can be considered to
be negligible with respect to the matter induced ones.
As a consequence, $\mit\Delta$
can be taken to be the most general $3\times 3$ unitary matrix,
that preserves appropriate consistent conditions: they assure 
the reality of the transition probability.
Taking into account these conditions,
$\mit\Delta$ can be parametrized in terms of two complex numbers
$u$ and $v$, such that $|u|^2+|v|^2=1$:
$$
{\mit\Delta}=\left(\matrix{
|u|^2-|v|^2 & \sqrt{2}\, \bar u v & \sqrt{2}\, u\bar v\cr
-\sqrt{2}\,\bar u \bar v & {\bar u}^2 & -{\bar v}^2 \cr
-\sqrt{2}\, u v & -v^2 & u^2}\right)\ .
\eqno(5.10)
$$
The explicit expression for the probability in (5.9) is now
rather involved; however, it simplifies when neglecting the fast 
oscillating terms:
$$
\Big\langle{\cal P}_{\nu_e\to\nu_\mu}(t)\Big\rangle=
{1\over 2}\Big\{ 1-e^{-2\alpha t}\ e^{\xi_f-\xi_i}\
\big(1-2\,|v|^2\big)\ \cos 2\phi_f\, \cos2\phi_i\Big\}\ .
\eqno(5.11)
$$

In practical applications, the interesting case occurs when the neutrinos
are generated in a medium with very large matter density, 
$e^{-\xi_i}\cos 2\phi_i\simeq -1$, while detected at a later time $t$ 
in vacuum, $\xi_f\simeq0$, $\phi_f=\theta$. In this case, one finds:
$$
\Big\langle{\cal P}_{\nu_e\to\nu_\mu}(t)\Big\rangle=
{1\over 2}\Big(1-e^{-2\alpha t}\Big)+e^{-2\alpha t}\Big[
\cos^2\theta-|v|^2\cos 2\theta\Big]\ .
\eqno(5.12)
$$
This is the most simple form that the transition probability
formula takes in presence of dissipative and matter effects:
with respect to the familiar expression, it contains exponential
damping factors. Taking into account that neutrinos are relativistic,
the flight time between emission and detection is with very good
approximation the same as the distance $\ell$ between source and
detector. One can then use (5.12) to derive a rough order of magnitude 
limits on the non-standard parameter $\alpha$. The best bounds are expected
from solar neutrinos, where $1/\ell$ can be as low as $10^{-27}\ {\rm GeV}$.

\vskip 2cm

{\bf 6. TRANSITION PROBABILITY IN VACUUM}
\medskip

One of the most interesting properties of the quantum dynamical semigroup
approach to neutrino propagation is the possibility of probing the
nature of the neutrinos by studying their oscillation pattern.
Indeed, the transition probability ${\cal P}_{\nu_e\to\nu_\mu}$
in (3.13) explicitly depends on the extra phase $\varphi$,
which can be non-vanishing for Majorana neutrinos.
In this section we shall discuss to what extent the phase $\varphi$ 
can be extracted from ${\cal P}_{\nu_e\to\nu_\mu}$.
With in mind possible applications to atmospheric neutrinos,
we shall limit our considerations to oscillations in vacuum,
and present various explicit formulas for the transition
probability in different approximations.

As a working assumption, let us first take the dissipative
parameters $c$ and $\beta$ in (3.5)
to be much smaller than the remaining constants.
To lowest order, a closed form for the entries of the evolution matrix
${\cal M}(t)$ in (3.8) can then be obtained. Using the general formula (3.13),
one finds:
$$
\eqalign{
{\cal P}_{\nu_e\to\nu_\mu}(t)={1\over 2}\Big\{
1-e^{-2\gamma t} \cos^2 2\theta &- e^{-(a+\alpha)t} \sin^2 2\theta
\Big[\cos (2\,\Omega\, t)\cr
&+ {|B|\over2\Omega}\
\sin(2\,\Omega\, t)\ \cos(\phi_B+2\varphi)\Big]\Big\}\ ,
}
\eqno(6.1)
$$
where $B=\alpha-a+2ib\equiv|B| e^{i\phi_B}$ and
$\Omega=\sqrt{\omega^2-|B|^2/4}$.
In this case the dependence on the phase $\varphi$ is very mild,
and can not be extracted by studying ${\cal P}_{\nu_e\to\nu_\mu}$ alone:
an independent determination of the combination $B$ is necessary.

The situation is even worse when $\gamma=\,0$; in this case, the
inequalities (2.3) automatically guarantee $c=\beta=\,0$
and further impose $b=\,0$, $a=\alpha$. In this case, $\varphi$
completely disappears from the expression of the transition probability,
that reduces to:[18]
$$
{\cal P}_{\nu_e\to\nu_\mu}(t)={1\over2}\sin^2 2\theta\,
\Big[1-e^{-2\alpha t}\cos(2\,\omega t)\Big]\ .
\eqno(6.2)
$$
In view of this, analysis of the experimental data based on (6.2)
along the lines of Ref.[32] is totally insensitive to $\varphi$;
a fit with more than one non-vanishing dissipative parameters
is in general needed, although this condition is certainly not enough,
as shown by (6.1).

In this respect, a more interesting situation occurs when $c=\,0$ and
$a=\alpha=\gamma$, as considered in the previous sections.
All the entries of the evolution matrix ${\cal M}(t)$ are now
non-vanishing, and the transition probability can be written as:
$$
\eqalign{
{\cal P}_{\nu_e\to\nu_\mu}(t)={1\over2}\bigg\{1&-e^{-2\alpha t}\bigg[
\bigg(1+{2\beta^2\over\Omega^2_0}\sin^2(\Omega_0 t)\bigg)\cos^2 2\theta\cr
&+\sin^2 2\theta\bigg(\cos(2\Omega_0 t)-
{2\beta^2\over\Omega^2_0}\sin^2(\Omega_0 t)\, \cos^2\varphi
-{b\over\Omega_0}\sin(2\Omega_0 t)\, \sin 2\varphi\bigg)\cr
&+{2\beta\over\Omega_0}\cos 4\theta\, \sin(\Omega_0 t)\bigg(
{b\over\Omega_0}\sin(\Omega_0 t)\, \cos\varphi-\cos(\Omega_0 t)\, \sin\varphi
\bigg)\bigg]\bigg\}\ ,
}
\eqno(6.3)
$$
where $\Omega_0=\sqrt{\omega^2-b^2-\beta^2}$ as before.
In this case, the dependence of ${\cal P}_{\nu_e\to\nu_\mu}(t)$ on time
is significally altered by the presence of a non-vanishing $\varphi$.

Atmospheric neutrino data are the most suitable for an experimental
study of (6.3), since in this case the time dependence can actually be probed.
Nevertheless, it should be stressed that the expression in (6.3) contains
four additional parameters besides the standard ones, $\omega$ and $\theta$,
so that the fitting procedure might turn out to be difficult in practice.
In order to simplify the analysis, one can further assume 
one of the two non-standard parameters $b$ or $\beta$ to be zero, but not both:
here again, when $\alpha$ is the only non-vanishing dissipative
parameter, the dependence on $\varphi$ in ${\cal P}_{\nu_e\to\nu_\mu}(t)$
disappears. Despite these difficulties, the amount of data on atmospheric
neutrinos is constantly growing, so that at least some information
on the presence of a non-vanishing $\varphi$, together with some of the
dissipative parameters, will surely be attainable in the near future.

\vskip 2cm

{\bf 7. DISCUSSION}
\medskip

The study of open systems by means of quantum dynamical semigroups
offers a physically consistent, general approach to the discussion of
phenomena leading to irreversibility and dissipation.
When this formalism is applied to the analysis of the propagation
of neutrinos, both in vacuum and in matter, it gives precise 
predictions on the pattern of oscillation phenomena:
the new, non-standard effects manifest themselves through a set of
phenomenological parameters, $\omega_1$, $\omega_2$, $\omega_3$
(the hamiltonian ones), $a$, $b$, $c$, $\alpha$, $\beta$, and $\gamma$
(the purely dissipative ones). Their presence allows oscillating
phenomena even for mass-degenerate neutrinos, accompanied by
possible $CP$-violating effects. Further, these predictions can be 
experimentally probed.
Indeed, fits of experimental data along the lines discussed in [32]
can be repeated for the more complete expressions of the
transition probability ${\cal P}_{\nu_e\to\nu_\mu}(t)$ presented
in the previous sections. 

For sure, the fitting procedure will be more
difficult and uncertain than in the standard case,
due to the presence of more unknown parameters.
Furthermore, one has to take into account that 
the effects induced by the presence of the new parameters
are expected to be small. By viewing the neutrinos as an open system
in weak interaction with an environment generated by a fundamental
``stringy'' dynamics, the effects of irreversibility and dissipation
can be roughly estimated to be proportional to the square of the average 
neutrino energy, divided by the characteristic energy scale of the environment.
Assimilating this scale to the Planck mass produces estimates of
order $10^{-27}\ {\rm GeV}$ for solar neutrinos, while larger
values are expected for more energetic neutrinos.
Despite these difficulties, a new generation of dedicated neutrino
experiments are presently collecting data or will shortly start 
construction, so that stringent bounds on the dissipative effects can
surely be expected in the future.

The above estimate on the magnitude of the non-standard, dissipative effects
is based on very general and physically motivated considerations
about open systems; therefore, it is rather robust and quite independent
from the details of the microscopic, fundamental dynamics responsible
for the interaction between the neutrino subsystem and the environment.
Nevertheless, it has been questioned on the basis of a formal similarity
of a particular, simplified version of the evolution equation (2.2)
with those describing the phenomenon of the so-called dynamical reduction
of the wave-packet.[48] The analogy is rather superficial: the physical
process leading to dissipation and the reduction process are quite 
distinct and act at different energy scales. Furthermore, as a more 
complete analysis would reveal, quantum dynamical semigroups generated
by equations of the form (2.2) are unable to properly describe
dynamical reduction processes.[9]

A different criticism on the use of quantum dynamical semigroups
for the description of dissipative effects advocates the use of non-linear
evolution equations.[49] Once more the general theory of open systems
offers a clarifying discussion on this point (for further technical
details, see the Appendix).

As pointed out in Sect.2, the dynamics of a small system $\cal S$
in interaction with a large environment $\cal E$ is in general very complex
and can not be described by means of evolution equations that are
linear in time: possible initial correlations and the continuos 
exchange of energy as well as entropy between $\cal S$ and $\cal E$
produces memory effects and non-linear phenomena.
Nevertheless, when the typical time scale in the evolution 
of the subsystem $\cal S$ is much larger than the characteristic time 
correlations in the environment, the subdynamics simplifies
and a mathematically precise description
in terms of quantum dynamical semigroups naturally emerges.[1-3, 21]

This limiting procedure is general and can be applied to all
physical situations for which the interaction between $\cal S$ and $\cal E$
is weak and for not too-short times, so that the non-linear disturbances
due to possible initial correlations have died out.[4]
These are precisely the conditions that are expected to be fulfilled in the
case of neutrino systems: the characteristic time correlations in the
environment, induced by the fundamental (gravitational or stringy)
dynamics, is certainly much smaller than the neutrino propagation time,
while the interaction between neutrinos and environment is for sure weak.

The description of neutrino propagation in terms of quantum dynamical 
semigroups automatically guarantees the fulfillment of basic physical
requirements, as forward in time composition, 
entropy increase (irreversibility)
and complete positivity. This is a clear advantage over alternative
formulations. Based on ideas originally presented in [23], generalized
dynamics for the neutrino system incorporating some of these
properties have been discussed before (see [50-52]). However,
those dynamics do not satisfy the condition of complete
positivity; as already mentioned, this could lead to serious inconsistencies
that can be avoided in all situations only by adopting evolutions
of the form (2.2).[41]

Non-linear dynamics in the description of the neutrino system naturally
emerge when the requirement of weak coupling between neutrinos and
environment is not satisfied. This typically happens in extreme
conditions, as those found in the core of a supernova or the early
universe; more conventional dissipative phenomena then arise due to
the scattering and absorption processes in the medium.[53-55] In order to
properly deal with these situations, a second-quantized, field-theoretical
formalism has been constructed, using specific effective interaction
hamiltonians as starting point. Although derived using techniques similar 
to the ones described before, the resulting kinetic evolution equations
are quite distinct from (2.2); they give rise to decoherence effects 
that modifies the pattern of neutrino oscillations in a very different
way with respect to the expressions discussed in the previous sections.
Nevertheless, also in this case the condition of complete positivity
needs to be satisfied for consistency, and this requirement might
produce further constraints on the modified dynamics.

Decoherence effects in neutrino physics have been further discussed
in connection with the uncertainties in the emission and detection processes.[56]
By smearing the familiar expression for the transition probability
over energy and time (or position) with an appropriate Gaussian distribution,
an exponential damping factor is generated, so that the resulting
expression for the averaged probability looks similar to
the one presented in (6.2). The analogy is once more only superficial,
since the transition probability (6.2) (and more generally the
expression in (3.13)) has an explicit time (position)
dependence that can not be reproduced via a Gaussian average.
Further, the physical mechanisms leading to the modified probability 
expressions are clearly different, the detector ``noise'' in one case, 
a fundamental dynamics in the other; in turn, this leads
to a different dependence of the damping factors
on the average neutrino energy.

As mentioned before, quantum dynamical semigroups can be employed to model
a large variety of physical situations. It is not a surprise
that they have been used to study the effects of density waves in the
propagation of neutrinos in fluctuating media, in particular, in the
interior of the sun; these phenomena are also described by equations
of the form (2.2) and induce modifications on the 
neutrino oscillation probabilities.[57] However,
it should be stressed that these density fluctuations
have their origin in the dynamics of the sun and operate at energy
scales quite different from the Planck mass;
therefore, they can be easily isolated from the dissipative effects
discussed before, that are not expected to be influenced by
long-range phenomena.

As a final remark, let us point out that several unconventional 
phenomena affecting neutrino propagation have been discussed in the
literature; they include: neutrino decay, flavour changing neutral currents,
violation of Lorentz invariance or the equivalent principle
({\it e.g.} see [58-64]).
All these phenomena lead to modifications of the standard
oscillation pattern; however, the resulting transition probability
${\cal P}_{\nu_e\to\nu_\mu}(t)$ has a dependence on time 
\vfill\eject\noindent
(or pathlength)
and neutrino energy that differ from the one discussed 
in the previous sections.

Indeed, the dependence of the observable ${\cal P}_{\nu_e\to\nu_\mu}(t)$
on the phenomenological parameters $a$, $b$, $c$, $\alpha$, $\beta$ 
and $\gamma$ is very distinctive of the presence of dissipative phenomena 
and can not be mimicked by other unconventional mechanisms. This
further strengthens the possibility of 
identifying the dissipative contributions
from the analysis of experimental data, quite independently from
other effects.

\vskip 2cm

{\bf APPENDIX: THE WEAK COUPLING LIMIT}
\medskip

For a neutrino system in interaction with an environment $\cal E$, the total
Hamiltonian can be decomposed as in (2.4):
$$
H_{\rm tot}=H\otimes {\bf 1} + {\bf 1}\otimes H_{\cal E} + 
g\,H'\ ,
\eqno(A.1)
$$
where, neglecting for simplicity matter effects, 
the system hamiltonian $H$ can be taken to be of the form
$$
H=\left(\matrix{E-\omega_0 & 0\cr
                   0 & E+\omega_0}\right)\ .
\eqno(A.2)
$$
Assuming no initial correlations between the two systems, the evolution
in time of the neutrino state $\rho(t)$ follows the general rule (2.5):
\null
$$
\rho\mapsto\rho(t)=
{\rm Tr}_{\cal E}\Big[ e^{-iH_{\rm tot} t}\,
\big(\rho\otimes\rho_{\cal E}\big)\, e^{iH_{\rm tot} t}\Big]\ ,
\eqno(A.3)
$$
\null
This evolution map is in general very complicated, developing
irreversibility and memory effects. However, it simplifies
when the interaction between the neutrino subsystem and the environment
is weak.

There are essentially two different ways of implementing 
in practice this \hbox{condition:[1-3]}
they correspond to the two ways of making the ratio $\tau/\tau_{\cal E}$
large; here, $\tau$ is the typical variation time of $\rho(t)$, 
while $\tau_{\cal E}$ represents the typical decay time of the 
correlations in the environment.
Only when $\tau\gg\tau_{\cal E}$, one expects the memory effects
in $(A.2)$ to be negligible, and a local in time
evolution for the state $\rho(t)$ to be valid.

When $\tau_{\cal E}$ becomes small, while $\tau$ remains finite, one speaks
of ``singular coupling limit'', since the typical time-correlations of
the environment approach a $\delta$-function. In the other case, 
it is $\tau$ that become large, while $\tau_{\cal E}$ remains finite:
one then works in the framework of the ``weak coupling limit''.
In practice, this is obtained by suitable rescaling the time variable, 
$t\rightarrow t/g^2$, and by sending the coupling constant $g$ 
to zero (van Hove limit).[1-3]

The choice between these two limiting procedures clearly depends
on specific physical considerations about the system under study.
In the case of the neutrino system, both limits appear physically acceptable.
Since the singular coupling limit has been presented elsewhere,[21] in the
following we shall concentrate on the discussion of the weak coupling limit.
In this case, following the steps presented in [21], from the finite
time evolution $(A.3)$ one can derive a differential equation for
$\rho(t)$ local in time; it is of the form $(2.2a)$,
$$
{\partial\rho(t)\over \partial t}= -i H\ \rho(t)+i\rho(t)\,
H + {\cal L}[\rho(t)]\ ,
\eqno(A.4)
$$
with
$$
{\cal L}[\rho]=-\lim_{T\to\infty}{1\over2T}\int_{-T}^T ds\ \int_0^\infty dt\ 
{\rm Tr}_{\cal E}\bigg\{ e^{iH_{\rm tot} s}\,\Big[ 
e^{iH_0 t}\, H'\, e^{-iH_0 t}\ , 
\big[H',\rho\otimes\rho_{\cal E}\big]\Big]e^{-iH_{\rm tot} s}\bigg\}\ ,
\eqno(A.5)
$$
where $H_0$ represents the limit of $H_{\rm tot}$ 
for a vanishing coupling constant $g$.

The general form of the additional term ${\cal L}[\rho]$ in $(A.5)$
does not actually depend very much on the details of the 
environment dynamics; an effective description that takes into account
its most fundamental characteristic properties is enough to allow an 
explicit evaluation of the integrals in $(A.5)$.
Following the idea that the dissipative effects are low energy phenomena
that originate from the fundamental gravitational or stringy
dynamics at some large  scale $M_F$, we shall model the environment
as a gas of quanta, obeying infinite statistics, in thermodynamic
equilibrium at inverse temperature $\beta_F=1/M_F$.%
\footnote{$^\dagger$}{For a motivation of this choice in terms of 
the dynamics of extended objects (D0-branes), see [65-67, 21].}

Further, taking into account that the interaction between the neutrino system
and the environment is weak, we shall assume the interaction hamiltonian
$H'$ to be linear both in the neutrino and the environment 
dynamical variables:
$$
H'=\sum_{\mu=0}^3\, \sigma_\mu \otimes B_\mu\ ;
\eqno(A.6)
$$
an explicit expression for the environmental operators $B_\mu$
will be discussed below.

In order to proceed further, it is convenient to introduce a 
spectral decomposition, and use the auxiliary matrices
$\sigma_\mu^{(\lambda)}$, $\lambda=-1, 0, 1$,[1]
$$
\sigma_\mu^{(0)}=P_1\, \sigma_\mu\, P_1 + P_2\, \sigma_\mu\, P_2\ ,\qquad
\sigma_\mu^{(+)}=P_1\, \sigma_\mu\, P_2\ ,\qquad
\sigma_\mu^{(-)}=P_2\, \sigma_\mu\, P_1\ ,
\eqno(A.7)
$$
with
$$
P_1=\left(\matrix{1 & 0\cr
                  0 & 0}\right)\ ,\qquad\quad
P_2=\left(\matrix{0 & 0\cr
                  0 & 1}\right)\ .
\eqno(A.8)
$$
Then, the limit in $(A.5)$ can be explicitly performed
and the result expressed in terms 
of the following $4\times 4$ hermitian matrices:
$$
\eqalignno{
&a_{\mu\nu}^{(\lambda)}=g^2 \int_{-\infty}^\infty dt\ e^{-2i\lambda\omega_0 t}\
\langle B_\mu(t)\, B_\nu\rangle\ , &(A.9a)\cr
&b_{\mu\nu}^{(\lambda)}=ig^2\bigg\{\int_0^\infty dt\ e^{-2i\lambda\omega_0 t}\
\langle B_\mu(t)\, B_\nu\rangle
- \int_0^\infty dt\ e^{2i\lambda\omega_0 t}\
\langle B_\mu\, B_\nu(t)\rangle\bigg\}\ , &(A.9b)}
$$
involving thermal correlations of the environment operators,
$$
\langle B_\mu(t)\, B_\nu\rangle={\rm Tr}_{\cal E}\Big[B_\mu(t)\,
B_\nu\ \rho_{\cal E}\Big]\ ,\qquad
\rho_{\cal E}={e^{-\beta_F H_{\cal E}}\over 
{\rm Tr}_{\cal E}\big(e^{-\beta_F H_{\cal E}}\big)}\ .
\eqno(A.10)
$$
Explicitly, one finds:
$$
{\cal L}[\rho]=i\big[\rho\ , \widetilde H\big]+L[\rho]\ ,
\eqno(A.11)
$$
where
$$
L[\rho]={1\over2}\,\sum_{\lambda\in\{0,\pm1\}}\,\bigg\{\sum_{i,j=1}^3 
a_{ij}^{(\lambda)}\, \Big[2\, \sigma_j^{(\lambda)}\rho\, \sigma_i^{(\lambda)} 
-\sigma_i^{(\lambda)}\sigma_j^{(\lambda)}\,\rho 
-\rho\,\sigma_i^{(\lambda)}\sigma_j^{(\lambda)}\Big]\bigg\}\ ,
\eqno(A.12)
$$
and
$$
\widetilde H={1\over2}\,\sum_{\lambda\in\{0,\pm1\}}\,\bigg\{\sum_{\mu,\nu=0}^3
b_{\mu\nu}^{(\lambda)}\, \sigma_\mu^{(\lambda)}\sigma_\nu^{(\lambda)}\bigg\}
+i\sum_{i=1}^3\big(a_{0i}^{(0)}-a_{i0}^{(0)}\big)\,\sigma_i\ .
\eqno(A.13)
$$

The first term in the r.h.s. of $(A.11)$ is of hamiltonian form.
This is a general feature of the reduced dynamics: even in absence of
an initial system dynamics, a non-trivial hamiltonian contribution is
always generated by the dissipative piece $(A.5)$. This mechanism has deep
consequences in neutrino physics: as remarked in the text, it allows 
oscillation phenomena even in case of initially massless neutrinos.

Further information on the coefficient matrices $a_{\mu\nu}^{(\lambda)}$
and $b_{\mu\nu}^{(\lambda)}$ in $(A.9)$ can be obtained 
by studying the behaviour of the environment correlations in $(A.10)$. 
The operators $B_\mu$ can be taken to be a general linear 
expression in the environment variables; these are the
creation, $A_a^\dagger(k)$, and annihilation, $A_a(k)$,
operators for the quanta representing the environment modes,
living in an abstract $n$-dimensional space:
$$
B_\mu(t)={1\over M_F^{(n-4)/2}}\sum_{a,b}
\int {d^{n-1}k\over [2(2\pi)^{n-1}\varepsilon(k)]^{1/2}}
\, f^a(k)\, \chi_\mu^{ab}\Big[A_b(k)\, e^{-i\varepsilon(k) t}
+A_b^\dagger(k)\, e^{i\varepsilon(k) t}\Big]\ .
\eqno(A.14)
$$
The coefficients $\chi_\mu^{ab}$ ``embed'' the $n$-dimensional environment modes 
into the effective two-dimensional
neutrino Hilbert space, while $f^a(k)$ are appropriate test functions 
necessary to make the
operator $B_\mu$ and its correlations well-defined; 
it can be taken to be of the form $(|k|/M_F)^{m/2}\, g^a(k)$,
for some positive integer $m$, with $g^a(k)$ of Gaussian form.
For sake of definiteness, in the following we shall use:
$g^a(k)=g^a(\Omega)\,e^{-\eta^2\varepsilon^2/2}$, with $g^a(\Omega)$
depending only on the angle variables and $\eta$ a real constant.
The function $\varepsilon(k)$ gives the dispersion relation
obeyed by the environment modes; for simplicity we shall adopt 
an ultrarelativistic law: $\varepsilon(k)=|k|\equiv\varepsilon$.
The powers of $M_F$, characterizing the energy scale of the environment, 
are necessary to give $B_\mu$ the right dimension of energy.

We assume an indefinite statistics for these modes, so that the creation
and annihilation operators obey generalized commutation relations:
$$
A_a(k)\, A_b^\dagger(k')- q\, A_b^\dagger(k')\, A_a(k)=
\delta_{ab}\ \delta^{(n-1)}(k-k')\ ;
\eqno(A.15)
$$
the real parameter $q$ determines the mode statistics:
the case $q=1$ corresponds to standard bosons,
while for $q=0$ one obtains the degenerate algebra discussed in [65-67],
in connection with D0-branes and black holes.
Without loss of generality, we shall assume $q< 1$.
Furthermore, the single-mode hamiltonian can be taken to be proportional 
to the corresponding number operator, 
so that the total environment hamiltonian $H_{\cal E}$ satisfies the relation
$$
[H_{\cal E}, A_a^\dagger(k)]=\varepsilon(k)\, A_a^\dagger(k)\ ,\qquad
[H_{\cal E}, A_a(k)]=-\varepsilon(k)\, A_a(k)\ ,
\eqno(A.16)
$$
implicit in the time dependence of $(A.14)$.

The thermal correlations involved in the definitions in $(A.9)$ can now
be readily computed. For instance, one explicitly gets:
$$
a_{ij}^{(\lambda)}
={1\over 2\, M_F^{m+n-4}}\int_{-\infty}^\infty \!\!\!
dt\, e^{-2i\lambda\omega_0 t}\
\int_0^\infty \!\! d\varepsilon\,
\varepsilon^{m+n-3}
\Big[X_{ij}(\varepsilon)\ e^{i\varepsilon t}+
X_{ji}(\varepsilon)\, e^{-i\varepsilon(t+i\beta_F)}\Big]\ 
{1\over e^{\beta_F\varepsilon}-q}\ ,
\eqno(A.17)
$$
where
$$
X_{ij}(\varepsilon)=\sum_{a,b,c}\bigg[\int{d\Omega_{n-1}\over(2\pi)^{n-1}}\
g^a(k)\ \chi_i^{ab}\, \chi_j^{cb}\ g^c(k)\bigg]
\equiv e^{-\eta^2\varepsilon^2}\ X_{ij}(0)\ ,
\eqno(A.18)
$$
involves the integration over the angle variables; notice that
$X_{ij}(\varepsilon)$ is a real, symmetric matrix.

This matrix is not generic: it turns out that in order to satisfy the
condition of entropy increase for finite $\beta_F$, 
$X_{ij}(\varepsilon)$ must vanish for $i,j=1,2$.
With this choice, one finds that the non-vanishing contributions
to $L[\rho]$ in $(A.12)$ can come only from the coefficient $a_{33}^{(0)}$,
provided $m=3-n$. Explicitly, one obtains:
$$
a_{33}^{(0)}={\pi\over 1-q}\ g^2\, M_F\, X_{33}(0)\ .
\eqno(A.19)
$$
The dimensionless coupling constant $g$, should be expressible 
in terms of the relevant energy scales, {\it i.e.} the average neutrino
energy $E$ and the mass $M_F$ characteristic of the environment.
Since $g$ is small, it must be at most of order $E/M_F$.
As a consequence, it turns out that the dissipative parameter
$a_{33}^{(0)}$ must scale as $E^2/M_F$. As mentioned in the text,
this is a general prediction of the open system approach to dissipation.

Using the expansion $\rho=\sum_\mu \rho_\mu\sigma_\mu/2$ as in the text,
one immediately finds that the dissipative contribution $L[\rho]$
in $(A.12)$ is of the form (3.3), with $a=\alpha=a_{33}^{(0)}$
and $b=c=\beta=\gamma=\,0$. In the weak coupling limit
a special form of the matrix (3.3) is then selected: it is
expressible in terms of only one non-standard parameter. This
situation does not hold any more in the case of the singular coupling
limit: in that case, all six parameters
$a$, $b$, $c$, $\alpha$, $\beta$, and $\gamma$ are in general non-vanishing
(for details, see [21]).

In a similar way, also the hamiltonian contribution in $(A.13)$ can be
explicitly computed. Taking into account the results obtained
in the evaluation of the coefficients $a_{ij}^{(\lambda)}$,
one sees that the $2\times 2$ matrix $\widetilde H$ becomes diagonal:
$$
\widetilde H=b_{03}^{(0)}\ \sigma_3\ ,
\eqno(A.20)
$$
where
$$
b_{03}^{(0)}=g^2\, M_F\, X_{03}(0)\ G\big(\beta_F/\eta\big)\ ,
\eqno(A.21)
$$
and
$$
G\big(\beta_F/\eta\big)=\int_0^\infty d t\
\int_0^\infty d\varepsilon\ \sin\varepsilon t\ e^{-\epsilon^2\eta^2}\
\bigg({1-e^{-\beta_F\, \varepsilon}\over 1-q e^{-\beta_F\, \varepsilon}}\bigg)\ .
\eqno(A.22)
$$
In the case of infinite statistics, $q=\,0$,
the function $G$ can be explicitly evaluated in terms of generalized
hypergeometric functions:
$$
G(x)={\sqrt{\pi}\over2}\, x\ 
{}_1F_1\bigg({1\over 2};{3\over2};{x^2\over4}\bigg)
-{x^2\over4}\ {}_2F_2\bigg(1,1;{3\over2},2;{x^2\over4}\bigg)\ .
\eqno(A.23)
$$

The operator in $(A.20)$ contributes via the parameter
$\omega_3=b_{03}^{(0)}$ to the effective hamiltonian $H_{\rm eff}$ 
in (3.1): even in absence of the standard piece $\omega_0$,
the quantity $\omega_3$ would still generate a level splitting between
the two neutrino mass eigenstates, making possible oscillation
phenomena.

As a further remark, notice that although both generated via the interaction
with the environment, the magnitude of the hamiltonian contribution
$\omega_3$ could differ from the dissipative one in $(A.19)$,
since their ratio involves the function $G$. Although in a different
context, this phenomenon has also been observed in [57].

\vfill\eject

\centerline{\bf REFERENCES}
\bigskip\medskip

\item{1.} R. Alicki and K. Lendi, {\it Quantum Dynamical Semigroups and 
Applications}, Lect. Notes Phys. {\bf 286}, (Springer-Verlag, Berlin, 1987)
\smallskip
\item{2.} V. Gorini, A. Frigerio, M. Verri, A. Kossakowski and
E.C.G. Surdarshan, Rep. Math. Phys. {\bf 13} (1978) 149 
\smallskip
\item{3.} H. Spohn, Rev. Mod. Phys. {\bf 52} (1980) 569
\smallskip
\item{4.} A. Royer, Phys. Rev. Lett. {\bf 77} (1996) 3272
\smallskip
\item{5.} C.W. Gardiner and P. Zoller,
{\it Quantum Noise}, 2nd. ed. (Springer, Berlin, 2000)
\smallskip
\item{6.} W.H. Louisell, {\it Quantum Statistical Properties of Radiation},
(Wiley, New York, 1973)
\smallskip
\item{7.} M.O. Scully and M.S. Zubairy, 
{\it Quantum Optics} (Cambridge University Press, Cambridge, 1997)
\smallskip
\item{8.} L. Fonda, G.C. Ghirardi and A. Rimini, Rep. Prog. Phys.
{\bf 41} (1978) 587 
\smallskip
\item{9.} G.C. Ghirardi, A. Rimini and T. Weber, Phys. Rev. D {\bf 34} (1986) 470
\smallskip
\item{10.} H. Nakazato, M. Namiki and S. Pascazio,
Int. J. Mod. Phys. {\bf B10} (1996) 247
\smallskip
\item{11.} F. Benatti and R. Floreanini, Phys. Lett. {\bf B428} (1998) 149
\smallskip
\item{12.} M.S. Marinov, JETP Lett. {\bf 15} (1972) 479; Sov. J. Nucl. Phys.
{\bf 19} (1974) 173; Nucl. Phys. {\bf B253} (1985) 609
\smallskip
\item{13.} J. Ellis, J.L. Lopez, N.E. Mavromatos 
and D.V. Nanopoulos, Phys. Rev. D {\bf 53} (1996) 3846
\smallskip
\item{14.} P. Huet and M.E. Peskin, Nucl. Phys. {\bf B434} (1995) 3
\smallskip
\item{15.} F. Benatti and R. Floreanini, Nucl. Phys. {\bf B488} (1997) 335
\smallskip
\item{16.} F. Benatti and R. Floreanini, Nucl. Phys. {\bf B511} (1998) 550
\smallskip
\item{17.} F. Benatti and R. Floreanini, Phys. Lett. {\bf B451} (1999) 422
\smallskip
\item{18.} F. Benatti and R. Floreanini, JHEP {\bf 02} (2000) 032
\smallskip
\item{19.} F. Benatti and R. Floreanini, Phys. Rev. D {\bf 62} (2000) 125009
\smallskip
\item{20.} J. Ellis, N.E. Mavromatos and D.V. Nanopoulos, Phys. Lett.
{\bf B293} (1992) 37; Int. J. Mod. Phys. {\bf A11} (1996) 1489
\smallskip
\item{21.} F. Benatti and R. Floreanini, Ann. of Phys. {\bf 273} (1999) 58
\smallskip
\item{22.} S. Hawking, Comm. Math. Phys. {\bf 87} (1983) 395; Phys. Rev. D
{\bf 37} (1988) 904; Phys. Rev. D {\bf 53} (1996) 3099;
S. Hawking and C. Hunter, Phys. Rev. D {\bf 59} (1999) 044025
\smallskip
\item{23.} J. Ellis, J.S. Hagelin, D.V. Nanopoulos and M. Srednicki,
Nucl. Phys. {\bf B241} (1984) 381; 
\smallskip
\item{24.} S. Coleman, Nucl. Phys. {\bf B307} (1988) 867
\smallskip
\item{25.} S.B. Giddings and A. Strominger, Nucl. Phys. {\bf B307} (1988) 854
\smallskip
\item{26.} M. Srednicki, Nucl. Phys. {\bf B410} (1993) 143
\smallskip
\item{27.} W.G. Unruh and R.M. Wald, Phys. Rev. D {\bf 52} (1995) 2176
\smallskip
\item{28.} L.J. Garay, Phys. Rev. Lett. {\bf 80} (1998) 2508;
Phys. Rev. D {\bf 58} (1998) 124015
\smallskip
\item{29.} F. Benatti and R. Floreanini, Phys. Lett. {\bf B401} (1997) 337
\smallskip
\item{30.} F. Benatti and R. Floreanini, Phys. Lett. {\bf B465} (1999) 260
\smallskip
\item{31.} F. Benatti and R. Floreanini, Irreversibility and dissipation
in neutral $B$-meson decays, Nucl. Phys. B, to appear, {\tt hep-ph/0103239}
\smallskip
\item{32.} E. Lisi, A. Marrone and D. Montanino,
Phys. Rev. Lett. {\bf 85} (2000) 1166
\smallskip
\item{33.} S.M. Bilenky and S.T. Petcov, Rev. Mod. Phys. 
{\bf 59} (1987) 671
\smallskip
\item{34.} C.W Kim and A. Pevsner, {\it Neutrinos in Physics and 
Astrophysics}, (Harwood Academic Press, 1993)
\smallskip
\item{35.} M. Fukugita and T. Yanagida, in {\it Physics and Astrophysics
of Neutrinos}, M. Fukugita and A. Suzuki, eds.,
(Springer, Tokyo, 1994)
\smallskip
\item{36.} R.N. Mohapatra and P.B. Pal, {\it Massive Neutrinos in Physics
and Astrophysics}, 2nd ed., (World Scientific, Singapore, 1999)
\smallskip
\item{37.} S.M. Bilenky, C. Giunti and W. Grimus, Prog. Part. Nucl. Phys.
{\bf 43} (1999) 1
\smallskip
\item{38.} J. Schechter and J.W.F. Valle, Phys Rev. D {\bf 22} (1980) 2227;
{\it ibid.} {bf 23} (1981) 1666
\smallskip
\item{39.} L.F. Li and F. Wilczek, Phys Rev. D {\bf 25} (1982) 143
\smallskip
\item{40.} B. Kayser, S.M. Bilenky, J. Hosek, S.T. Petcov, 
Phys. Lett. {\bf B94} (1989) 495; M. Doi, T. Kotani, H. Nishiura,
K. Okuda and E. Takasugi, Phys. Lett. {\bf B102} (1981) 323;
P. Langacker {\it et al.} Nucl. Phys. {\bf B282} (1987) 589
\smallskip
\item{41.} F. Benatti and R. Floreanini,
Mod. Phys. Lett. {\bf A12} (1997) 1465; 
Banach Center Publications, {\bf 43} (1998) 71; 
Phys. Lett. {\bf B468} (1999) 287; On the weak-coupling limit and complete
positivity, Chaos Sol. Frac., to appear
\smallskip
\item{42.} T.K. Kuo and J. Pantaleone, Rev. Mod. Phys. {\bf 61} (1989) 937
\smallskip
\item{43.} S.P. Mikheyev and A.Yu Smirnov, Sov. J. Nucl. Phys.
{\bf 42} (1985) 913; Nuovo Cim. {\bf 9C} (1986) 17
\smallskip
\item{44.} L. Wolfenstein, Phys. Rev. D {\bf 17} (1978) 2369;
{\it ibid.} {\bf 20} (1979) 2634
\smallskip
\item{45.} For instance, see: M. Artin, {\it Algebra}, (Prentice Hall,
Englewood Cliffs (NJ), 1991)
\smallskip
\item{46.} K. Lendi, J. Phys. {\bf A 20} (1987) 13
\smallskip
\item{47.} A.M. Gago, E.M. Santos, W.J.C. Teves, R. Zukanovich Funchal,
Quantum dissipative effects and neutrinos: current constraints and
future perspectives, {\tt hep-ph/0009222}
\smallskip
\item{48.} S. Adler, Phys. Rev. D {\bf 62} (2000) 117901
\smallskip
\item{49.} J. Ellis, N.E. Mavromatos and D.V. Nanopoulos,
Phys. Rev. D {\bf 63} (2001) 024024
\smallskip
\item{50.} Y. Liu, L. Hu and M.-L. Ge, Phys. Rev. D {\bf 56} (1997) 6648
\smallskip
\item{51.} C.-H. Chang, W.-S. Dai, X.-Q. Li, Y. Liu, F.-C. Ma and
Z. Tao, Phys. Rev. D {\bf 60} (1999) 033006
\smallskip
\item{52.} C.P. Sun and D.L. Zhou, Quantum decoherence effect and
neutrino oscillations, {\tt hep-ph/9808334}
\smallskip
\item{53.} A.D. Dolgov, Sov. J. Nucl. Phys. {\bf 33} (1981) 700
\smallskip
\item{54.} G. Raffelt, G. Sigl and L. Stodolsky, Phys. Rev. Lett.
{\bf 70} (1993) 2363; G. Sigl and G. Raffelt, Nucl. Phys.
{\bf B406} (1993) 423; G. Raffelt, {\it Stars as Laboratories
for Fundamental Physics}, (Chicago University Press, Chicago, 1996)
\smallskip
\item{55.} M. Prakash, J.M. Lattimer, R.F. Sawyer and R.R. Volkas,
Neutrinos in dense matter, early universe, supernovae, neutron stars,
{\tt atro-ph/0103095}
\smallskip
\item{56.} T. Ohlsson, Equivalence between Gaussian averaged neutrino
oscillations and neutrino decoherence, {\tt hep-ph/0012272}
\smallskip
\item{57.} C.P. Burgess and D. Michaud, Ann. of Phys. {\bf 256} (1997) 1;
P. Bamert, C.P. Burgess and D. Michaud, Nucl. Phys. {\bf B513} (1998) 319
\smallskip
\item{58.} P. Lipari and M. Lusignoli, Phys. Rev D {\bf 60} (1999) 013003
\smallskip
\item{59.} G.L. Fogli, E. Lisi, A. Marrone and G. Scioscia,
Phys. Rev. D {\bf 60} (1999) 053006
\smallskip
\item{60.} J. Alfaro, H.A. Morales-Tecotl and L.F. Urrutia,
Phys. Rev. Lett. {\bf 84} (2000) 2318
\smallskip
\item{61.} C.N. Leung, Neutrino tests of general and special relativity,
{\tt hep-ph/0002073}
\smallskip
\item{62.} H.V. Klapdor-Kleingrothaus, H. Pas and U. Sarkar,
Effects of quantum space time foam in the neutrino sector,
{\tt hep-ph/0004123}
\smallskip
\item{63.} V. Barger, S. Pakvasa, T.J. Weiler and K. Whisnant,
$CPT$-odd resonances in neutrino oscillations, {\tt hep-ph/0005197}
\smallskip
\item{64.} M. Lindner, T. Ohlsson and W. Winter, A combined treatment of
neutrino decay and neutrino oscillations, {\tt hep-ph/0103170}
\smallskip
\item{65.} A. Strominger, Phys. Rev. Lett. {\bf 71} (1993) 3397
\smallskip
\item{66.} I.V. Volovich, D-branes, black holes and $SU(\infty)$ gauge theory,
{\tt hep-th/9608137}
\smallskip
\item{67.} D. Minic, Infinite statistics and black holes in Matrix theory,
Pennsylvania University preprint, 1997, {\tt hep-th/9712202}

\bye